\documentstyle[sprocl,psfig]{article}

\def\be{\begin{equation}}
\def\ee{\end{equation}}
\def\bea{\begin{eqnarray}}
\def\eea{\end{eqnarray}}
\def\Pom{{\small I\!P}}

\def\gsim{\mathrel{\rlap{\lower4pt\hbox{\hskip1pt$\sim$}}
    \raise1pt\hbox{$>$}}}         
\def\lsim{\mathrel{\rlap{\lower4pt\hbox{\hskip1pt$\sim$}}
    \raise1pt\hbox{$<$}}}         

\begin{document}

\title{SUMMARY OF THE DIFFRACTIVE WORKING GROUP AT DIS98}

\author{A. Goussiou}

\address{Department of Physics and Astronomy \\
State University of New York at Stony Brook \\
Stony Brook, New York 11794, U.S.A. \\
E-mail: goussiou@fnal.gov} 

\author{M. McDermott}

\address{Department of Physics and Astronomy, Brunswick St, \\
 University of Manchester, M13 9PL, England \\
E-mail: mm@a13.ph.man.ac.uk}  

\author{N.N. Nikolaev}

\address{Institut f. Kernphysik, Forschungszentrum J\"ulich, D-52425 J\"ulich,
Germany
\& L.D.Landau Institute for Theoretical Physics, Kosygina 2, 117334 Moscow,
Russia \\
E-mail: N.Nikolaev@fz-juelich.de}

\author{R. Roosen}

\address{I.I.H.E, Vrije Universiteit Brussel, \\
         Pleinlaan 2,1050 Brussels, Belgium  \\
         E-mail: roosen@hep.iihe.ac.be}

\author{K. Piotrzkowski}

\address{DESY,  Notkestrasse 85, D-22607 Hamburg
  Germany\\ 
E-mail: krzysztof.piotrzkowski@desy.de}


\maketitle\abstracts{Recent experimental and theoretical developments
in the understanding of high energy diffraction, presented in the
working group on diffraction at DIS98 in Brussels, are summarised.
A template, giving the definition of the most commonly used kinematical
variables in diffraction, which was provided in the working group
sessions, is reproduced as a appendix. References to original 
papers may be found within the individual contributions.}

\section{Introduction}
\subsection{Why is diffraction interesting ?}

Diffraction combines aspects of particle and wave-like nature of
high energy scattering and straddles the interface between
short and long distance domains of the strong interaction.
Apart from being a very interesting problem in its own right,
it is also a useful place to try to understand the transition between
reliable, perturbative QCD calculations and the remarkably successful
strong interaction phenomenology of Regge theory. Such an understanding
is clearly necessary if one is to ever understand the most
difficult and important problem in strong interaction physics: confinement.

Given that a fundamental understanding of this transition is still lacking,
 progress in this area may be characterised as follows:
one investigates what can be understood in the regime of pQCD, 
extrapolates this hard QCD wisdom into the soft domain,
often using experience and intuition gained from soft physics phenomenology,
and then confronts the results with the data.
This necessarily leads to a strong positive feedback between
new experimental results in diffraction and the development and
refinement of the phenomenological models.
A great deal of experimental and theoretical progress
has been made since DIS97 which we wish to summarise here.

\subsection{Physical picture of high energy diffraction}

Peschanski~\cite{peschanski} reminded us of a very 
simple physical picture 
for diffraction based on an optical model,
developed many years ago. Imagine two
hadrons scattering at very high energies. Quantum mechanics tell us that
each hadron is a complicated evolving superposition of virtual states
(at long distances one can
think of the proton emitting and reabsorbing pions, etc;
at short distances
one imagines fluctuations of the partonic structure due QCD radiation).
Lorentz contraction of
these ultra-relativistic systems ensures that this superposition is
essentially frozen on the
`snapshot' timescale of the interaction of the two systems, thus
 each component constitutes an eigenstate of the interaction.
Distinct eigenstates will suffer different levels of 
attenuation in the nuclear medium of the
opposite hadron according to their physical
characteristics (number of constituents/partons, transverse size etc).
As a result the scattered state is different from the beam state and,
in addition to elastic scattering,
new {\it exclusive and continuum  states}
are `diffracted into existence' by the interaction.
In such diffractive production there is a large rapidity gap (LRG) between the 
beam excitation and the target recoil and the energy dependence is similar to
that of elastic scattering: in the Regge terminology
both elastic scattering and diffractive excitation are governed by Pomeron
(vacuum) exchange in the $t$-channel.

If the Pomeron is an isolated Regge pole with the trajectory
$j=\alpha_{\Pom}(t)\approx \alpha_{\Pom}(0)+\alpha_{\Pom}'t$, then
diffractive amplitudes are expected to behave as
$x_{\Pom}^{-\alpha_{\Pom}(t)}$. It would certainly be interesting to relate this
rise in energy with that seen in the 
proton structure function at small-$x$. The Regge phenomenology
of hadronic total cross sections provides a useful reference
value
\be
\alpha_{\Pom}^{soft}(0)\approx 1.09.
\label{softPom}
\ee
The QCD vacuum singularity seems to be more complex than an isolated pole
with an effective Pomeron trajectory that changes with the hardness of
the process. Any value larger than that given in
Eq.(\ref{softPom}) may be considered
as  evidence for a contribution from hard scattering.

\section{Diffraction in deep inelastic scattering}

\subsection{Inclusive data}

The two measurements reported on by Kowalski \& Lindemann for ZEUS~\cite{kowalski}
and by Nicholls for H1~\cite{nicholls}, regarding the diffractive structure 
function $F_{2}^{D(3)}$, are based on two different methods and cover
different kinematic ranges. The ZEUS data are analysed in the kinematic 
range $7\leq Q^2\leq 140~GeV^2$ and $M_X\leq 15~GeV$ and are based on the 1994
data. The diffractive data are obtained from an excess of events over 
the extrapolated invariant mass distribution at large $M_X$ in 
generic DIS events. 
The H1 diffractive data are obtained from events
with a large rapidity gap in the forward direction. The H1 
diffractive structure function analysis is based on the 1994 data,  
complemented by the new 1995 shifted-vertex data which 
covers $0.4\leq Q^2 \leq 5 GeV^2, 0.001\leq \beta \leq 0.65$,
thereby  extending  the 1994 measurements to lower $Q^2$, $\beta$ and $x_\Pom$.
A comparison of  the inclusive $x_\Pom .F_{2}^{D(3)}$ data of H1, 
ZEUS and of the ZEUS Leading Proton Spectrometer (LPS)
data shows that there is broad 
agreement, although in the low-$Q^2$ bins differences are observed.
Phenomenological Regge model fits as used previously, based on a Pomeron and
Reggeon trajectory describe the H1 data well. The 
intercept of the trajectories 
are consistent with the earlier published values and, given the large errors,  
no evidence is found for a possible dependence of the Pomeron 
intercept on  $Q^2$. The scaling violations, observed earlier 
in $x_{\Pom}.F_{2}^{D(3)}(x_{\Pom}=0.005)$  of the 1994
data at higher $Q^2$, are reconfirmed by the new data at lower $Q^2$. 
The analysis of the data in terms of parton distribution functions 
subjected to a NLO DGLAP evolution, again
indicate that the gluonic content of the Pomeron is of the order of (80-90)\%
with a gluon distribution which is large at $\beta\sim 1$, 
in contrast to a much softer gluon content in the proton.

From the  ZEUS analysis 
an intercept of $\alpha_\Pom(0)$ is deduced which agrees with the H1 value 
obtained from the phenomenological fits of the 1994 data.
\begin{eqnarray}
\alpha_\Pom(0) &=& 1.16 \pm 0.01~(\mbox{stat}) \pm 0.02~(\mbox{sys})   ~~(\mbox{ZEUS})  \\
\alpha_\Pom(0) &=& 1.203\pm 0.020~(\mbox{stat}) \pm 0.013~(\mbox{sys})  ~~(\mbox{H1-`94 data})
\end{eqnarray}
This intercept is clearly larger than that of Eq.(\ref{softPom}), 
whereas the Reggeon intercept,
obtained by H1, is close to $\alpha_R(0)\approx 0.5$ of standard Regge theory.
The ZEUS results for  $x_{\Pom} \cdot F_{2}^{D(2)}$ indicate
a weak $\beta$-dependence and are, within errors, consistent with 
scaling.

\subsection{Models of diffractive DIS}

Diffraction occurs in the small-$x$ regime of DIS
corresponding to the high energy (Regge)
limit of the $\gamma^{*} p$ sub-process ($W^2 \gg Q^2, M_p^2$).
In the realm of QCD, the multi-parton Fock states of the photon are the
natural diffraction eigenstates.
At lowest order in $\alpha_s$ these are
$q \bar q$ pairs: colour dipoles characterised by transverse size,
or equivalently impact parameter, and momentum-fraction sharing, z.
The dipole scattering amplitudes are proportional
to the transverse area occupied by the dipole,
hence it is large size configurations
which are primarily responsible for diffraction (they also turn out to
be asymmetric configurations $z \ll 1$, or $1-z \ll 1$).
As the transverse size, or `scanning radius', of the interacting
dipoles is decreased one expects a transition from soft to hard
diffraction.
Genovese~\cite{genovese} reviewed major applications of the colour dipole
picture to diffractive DIS.

Using the predictions of a dipole model approach, based on leading-log 
BFKL dynamics and the large Nc approximation, Royon~\cite{royon} presented 
fits to the H1 diffractive data, as well as to the $F_2$ data.
Both the proton and virtual photon are treated as a superposition of dipoles
and both single and double diffraction are included in terms of
elastic onium-proton scattering and a sum of inelastic
dipole-dipole scattering. The gross features of the experimental data can
be reproduced with relative few parameters,
but the applicability of these approximations
to diffractive DIS is certainly questionable,
due to the large soft contribution to diffractive DIS.

Kopeliovich~\cite{kopeliovich} discussed how Drell-Yan
production, which is usually treated as a $q\bar{q} \rightarrow
\gamma^{*}$ annihilation, can be reformulated in the colour
dipole picture as a sort of diffractive excitation of a
Fock state of the projectile which contains the $\gamma^{*}$ as
a constituent. In this way, the similarity between Drell-Yan
and DIS processes becomes apparent.

It is useful to focus attention on those (relative rare)
diffractive processes which also contain a hard scale, in addition to $Q^2$,
such as a heavy quark mass or high-$p_t$ pair, to be really 
sure that we can trust our perturbative calculations. 
The knowledge gained can then be used to
try to build an understanding of the wider picture of diffractive processes.
The QCD model for
this `hard' diffraction
is the exchange of two interacting gluons in a colour-singlet configuration
in the $t$-channel, which dominates the QCD evolution of the
proton sea structure function at small $x$.
The diffractive amplitudes
are related to the target gluon structure function, 
$G(x_{\Pom},\overline{Q}^{2})$,
at a process-dependent hardness scale $\overline{Q}^{2}$.
As Sch\"afer~\cite{schaefer} has discussed, one can
similarly view Reggeons as the exchange of colourless
$q\bar{q}$ pairs in the $t$-channel and relate them to the valence
quark component of the proton structure function.

In the case of hard diffraction,
one calculates the characteristics  of
diffractive scattering of different dipoles
($q {\bar q},q {\bar q} g$, etc) as a
function of $\beta$ and $Q^2$, at fixed $x_\Pom$, 
largely by knowing the wavefunctions of the  longitudinal
and transverse photons.
An important finding~\cite{genovese} from these studies is
a strong process dependence of the hardness scale $\overline{Q}^{2}$ and  
a lack of overall Regge factorization into an $x_{\Pom}$-dependent
flux and a $(\beta,Q^{2})$-dependent structure function of the
Pomeron, apart from the region of small $\beta$. Furthermore, the values of
$\overline{Q}^{2}$ and consequently the 
$x_{\Pom}$- and $\beta$-dependence are strongly
affected by the presence of the additional hard scale (mass of
the heavy flavour, $k_{\perp}^{2}$ of jets, etc).
For large-$\beta$ the fluctuation of
the longitudinal photon into a $q \bar q $ is expected
to dominate, even though formally it is higher twist, which is
a situation unprecedented in inclusive DIS.
Pronyaev~\cite{pronyaev} discussed how this pQCD result, 
which is caused by the interference of diffraction
of longitudinal and transverse photons,  can be
tested by measuring the azimuthal correlation of the $(e,e')$ and $(p,p')$
scattering planes.
Moving to intermediate $\beta$, the
$q {\bar q}$ from the transverse photon becomes increasingly important,
whereas for very large masses at small
$\beta$ ($M_X^2 \gg Q^2$) it is likely that higher Fock states, e.g.
$q {\bar q} g$ will dominate as the phase space opens up to
allow additional radiation.

A recent reanalysis, in this context,
involving a sensible extrapolation  of this wisdom
to the whole of the region covered by the diffractive cross section 
measurement was presented by W\"usthoff~\cite{wusthoff}.
The diffractive structure function,  parametrised as
\[
x_\Pom F_{2}^{D(3)}(\beta,x_\Pom,Q^2)
= c_t.F^{T}_{q\bar q}+c_L.F^{L}_{q\bar q}+c_g.F^{T}_{q\bar qg}
\]
is fitted to the diffractive data as a 9 parameter function where
$F^{T}_{q\bar q},F^{L}_{q\bar q}$ and $F^{T}_{q\bar qg}$ stand for the
longitudinal and transverse contributions of the photon Fock
states $q\bar q$ and $q\bar q g$.
This expression has been fitted to the H1 and ZEUS diffractive data and
is able to describe both data sets well. In the fit to the H1 data,
which covers a wider kinematic range, a contribution of secondary
Reggeons is also taken into account.

It turns out that the present data do not allow
the $\beta$-dependence of the quark and gluon distributions in the
Pomeron to be fixed uniquely. In one fit the
$F^{T}_{q\bar qg}$ contribution dominates
the diffractive structure function at low $\beta$ and
falls steeply as  $\beta\rightarrow 1$, as preferred by some theorists.
At medium and large $\beta$, respectively, the transverse
$q \bar q$ term and the longitudinal $q\bar q$ higher twist term are dominant.
The observed scaling violations can then be ascribed to the 
(Regge-factorization breaking !)
$Q^2$-dependence 
of the $x_\Pom$ exponent in the
transverse $q \bar q$ term.
In addition, the H1 data also allow  a second solution
in which the $\beta$-dependence of the $F^{T}_{q\bar qg}$ contribution is much
harder and dominates over much of the $\beta, Q^2$ range of the data.

\subsection{Are diffractive events universal ?}

At the moment the Regge (Ingelman-Schlein) factorization ansatz
remains the only tool to relate diffractive cross sections in DIS
and hadronic collisions. It is of limited applicability and a 
better understanding of the consequences of the 
process-dependent hardness scale
$\overline{Q}^{2}$ is needed. The related assumption of hard
scattering factorization 
when all diffractive processes are described in terms
of universal diffractive parton densities
is also questionable because of
strong absorption, which goes under the name of Bjorken's gap
survival probability,  in hadronic hard diffraction.

Whitmore~\cite{whitmore} has reported a detailed evaluation of
hard diffraction cross sections at the Tevatron (dijets, W-production)
based on fits to diffractive DIS but restricted to a subset of
the inclusive H1 and ZEUS DIS data, as
well as the ZEUS diffractive jet photoproduction data. 
He concludes from the comparison with the D0 and CDF data
that there is a breaking of factorisation
for the Tevatron ($p\bar p$) results, in line with Bjorken's
small gap survival probability. 
A comparison of predictions for diffractive dijets and
diffractive W production 
will eventually help 
to pin down the relative quark-gluon contents of the Pomeron. 

Schaefer~\cite{schaefer} has reminded us that nuclear shadowing is a
yet another observable which is calculable in terms of the
diffractive structure function. The NMC data on scaling violations
in nuclear structure functions are so accurate that one can
evaluate nuclear modifications of the gluon structure functions from
the DGLAP evolution analysis
reliably.
For the deuteron these results imply $\sim 3$\% shadowing at $ x\sim 10^{-4}$,
in close similarity to the observed shadowing of the sea.
Schaefer finds that if the gluons in the Pomeron
carry about the same momentum as quarks and antiquarks,
$\langle x_{g}\rangle \sim \langle x_{q,\bar{q}}\rangle$,
then nuclear shadowing of gluons in the deuteron will be
$\sim 3$\% at $x\sim 10^{-4}$, whereas an unacceptably large
10-15\% shadowing is found using the 
super-hard gluonic Pomeron advocated
by H1. This apparent contradiction needs to be understood and resolved.

\section{Aspects of exclusive production}

\subsection{Vector meson production: data}

The study of diffractive vector meson production at HERA remains a very
active research field 
and is an ideal place to 
study the transition between hard and soft diffraction. 
The former is characterised 
by stronger energy rises, broader diffractive peaks 
and considerably less shrinkage than the latter.
The first observation of the  photoproduction of the $\Upsilon$-family
was reported for ZEUS by Bruni~\cite{bruni}. 
A cross-section ($\sigma_{\gamma p\rightarrow 
\Upsilon(nS) p} * BR(\Upsilon(nS) 
\rightarrow \mu^+ \mu^-) $ for $n=1,2,3$) of $\approx 15 \mbox{pb} $  
has been extracted using the full
95-97 data statistics, the branching ratios  
$\Upsilon(nS)\rightarrow\mu^+\mu^-$, 
an estimate of proton-dissociation, and an
assumption of the same relative contributions of 
$\Upsilon(1S), \Upsilon^{'} (2S)$ and $\Upsilon^{''} (3S)$ states as measured 
at the  Tevatron. 
In spite of the large scale given by the $\Upsilon$ mass 
(which should make pQCD prediction reliable),
even taking into account the large uncertainties due
to the choice of the gluon density, the scale it is sampled at, 
and the choice of light-cone wavefunction of the vector meson,
it turns out that the predictions of a pQCD two-gluon exchange model,
which successfully describes the $J/\psi$ production, are
about an order of magnitude below the measured cross section. 
Clearly further developments in both the experimental measurement 
(reduction of large errors) and theoretical 
understanding are urgently required.

Monteiro~\cite{monteiro} reported for ZEUS on exclusive and proton-dissociative
photoproduction of $\rho^0, \phi$ and $J/\psi$  mesons at
$W\approx100$~GeV and  $0<|t|<4$~GeV$^2$.
Using the Regge formalism and the measured elastic cross-sections, as well as
the low-$W$ data and other HERA measurements at low $|t|$,
the exchanged Pomeron trajectory could
be directly determined up to $|t|\approx 1$~GeV$^2$.
For the $\rho^0$ and $\phi$ production the nominal `soft' (linear) trajectory
has been measured
with an intercept compatible with 
Eq.(\ref{softPom}). The slope of the trajectory is non-zero,
but it is significantly smaller than $0.25$~GeV$^{-2}$.
In contrast, for $J/\psi$ exclusive production, 
the corresponding trajectory has a much higher 
intercept {\it and} its slope is 
small, compatible with zero,
indicative  of a small transverse size and 
a `hard' diffractive mechanism.

Thompson~\cite{thompson} reported studies of diffractive $J/\psi$ 
photo- and electroproduction and also photoproduction at high-$|t|$ for H1.
For  $|t| > 1 $~GeV the measured cross-section for proton-dissociative
diffraction can be successfully described 
by a model based on LO BFKL  (see also Sec.(7)). 
New measurements of exclusive electroproduction of the 
small-size $J/\psi$ confirm
the strongly rising $W$-dependence already seen in photoproduction. 
The ratio of $\psi(2S)$ to $J/\psi$ production is found 
to increase from about 
$0.15$ in photoproduction to about $0.5$ at  
$Q^2 \approx 15$~GeV$^2$, with large errors. 
This reflects the increase in hardness of the production scale of
$\psi(2S)$ with $Q^2$, which is similar in size to the pion,
and may reveal important information about 
the light-cone wave-functions of these heavy vector mesons.

The diffractive electroproduction of $\rho^0$ mesons was reported by 
Clerbaux~\cite{clerbaux} for H1, Tytgat~\cite{tytgat} for HERMES and 
Kananov~\cite{kananov} for ZEUS (also $\phi$ mesons). 
All three experiments measured the ratio
$R=\sigma_L/\sigma_T$ 
from photoproduction up to large-$Q^2$ production.
The $Q^2$-dependence of $R$ is consistent with a linear increase
up to $Q^2 \approx 0.5$~GeV$^2$, beyond which this strong
increase becomes significantly weaker. 
The quantitative description of this behaviour, which is now well
established experimentally, is a challenge to
the pQCD based models.


Fredj~\cite{fredj} reported on an interesting 
contribution to diffractive physics from the L3 experiment at
LEP - the measurement of the 
$\gamma\gamma$ total cross section extending the energy
range up to $W_{\gamma\gamma}\approx$ 130 GeV. Fits 
to $W_{\gamma\gamma}^{2(\alpha_{\Pom}(0)-1)}$ give effective 
Pomeron intercepts, depending on the unfolding method,
of $\alpha_{\Pom}(0)\approx 1.16\pm 0.03$ (PHOJET Monto Carlo) and 
$\alpha_{\Pom}(0)\approx 1.14\pm 0.02$ (PYTHIA Monte Carlo), 
in excess of the value in Eq.(\ref{softPom}).

\subsection{Vector meson production : theory}

Zoller~\cite{zoller}  presented results on expectations for the forward
diffractive slope, $B_D$,  within the framework of the gBFKL dipole model.
Three components can be identified coming from the proton,
the evolution and from the scattering dipole.
As $Q^2$ increases the dipoles get smaller and
the latter makes a smaller and smaller contribution
to $B_D$,  supporting the well-established notion
that the geometrical size of the scattering
objects determines $B_D$.
Unfortunately the current experimental errors from HERA are
too big to observe this $Q^2$-dependence in $J/\psi$ production yet,
but for lighter vector mesons it has been well established
experimentally. An approximate flavour independence is
observed in the variable $Q^2 + M_V^2$.

The analyses~\cite{clerbaux,tytgat,kananov} 
of the vector meson production data,  
under the assumption of the $s$-channel helicity conservation, give a value for   
$R = \sigma_L/\sigma_T$ which tends to saturate at large $Q^2$,
whereas the theoretical estimates predict a steady rise,
albeit slower than linear with $Q^{2}$. In the framework of
the two gluon exchange model of Low-Nussinov, Royen~\cite{royen} 
discussed the sensitivity  of $R$
to modifications of 
the wavefunction of the vector meson with the conclusion that
the Fermi motion effects in the wavefunction  can tame the growth
of $R$
without spoiling other predictions, in the kinematic region defined by the data.

\subsection{Off-diagonal kinematics}

A particularly active area of research at present concerns non-diagonal
or off-forward parton distributions which arise in exclusive
diffractive processes such as heavy vector meson production,
deeply virtual Compton scattering (DVCS) and photoproduction of dijets.

Conventional parton distributions involve
products of operators sandwiched between identical hadronic states
(e.g {\it incoming} and {\it outgoing} protons in the same quantum state).
The finite momentum transfer to the proton, in non-diagonal kinematics,
means that the outgoing hadron (even if it is a proton) is in a
different quantum state.
This leads to universal distributions which are given by the quantum-mechanical
interference between states
characterised by the difference in the momentum fractions
carried by the outgoing ($x_1$) and returning ($x_2$) partons,
$x_{\Pom} = x_1 - x_2$,
and as such probe new non-perturbative
information about the proton. 
A renormalization group analysis of the operators leads
to evolution equations, dependent on $x_{\Pom}$,
which are known only to leading-log in $Q^2$, at present.
For $x_2 > 0$ they reduce to the
DGLAP equations in the limit $x_{\Pom} \rightarrow 0$.
For  $x_2 < 0$ they obey ERBL equations for the distribution amplitudes.
In the  leading $\ln(1/x)$ approximation, at small-$x$,
the non-diagonal distributions coincide with the conventional (diagonal)
gluon distributions.

Golec-Biernat~\cite{golec} presented interesting
results on diffractive dijet production which showed that
the next-to-leading $\ln(1/x)$ corrections for the non-diagonality 
of the process leads to a marked enhancement of jet production.
It would be interesting to see the impact of other non-leading
corrections on this finding.

Strikman~\cite{strikman} presented an analysis of the related 
off-diagonal DVCS, and pointed out the feasibility
of measuring the real part of the DVCS amplitude at HERA, which,
via dispersion relations, constrains the behaviour of the imaginary
part of the DVCS amplitude, 
and by extension $F_2$,
at smaller $x$-values  than those in the HERA kinematic range.

In conventional definitions of parton distributions,
after factorization into hard and soft physics,
one exploits the optical theorem and treats the `soft blob'
(and the hard blob) as though it were on mass-shell (as a result of the cut).
By considering the singularity structure of the four-point Green's function
for the soft blob in the non-diagonal case Diehl~\cite{diehl}
has shown that one may treat the soft part of the diagram
{\it as though it had been cut} and explains
some of the important physical implications of this result.

\section{Diffractive final states}

Various aspects of diffractive final states have been reported by 
Buniatian~\cite{bunatian} and Waugh~\cite{waugh} for 
H1 (energy flow, seagull plot, average charged particle 
multiplicities, mean 
multiplicities in the forward/backward hemispheres) and 
Wichmann~\cite{wichmann} for ZEUS 
(thrust and sphericity analysis) collaborations. 
The global features of diffractive final states, i.e. the rapidity 
and transverse momentum distributions at small $k_{\perp}$,
mean multiplicities and multiplicity distributions, the seagull plot, 
are similar to those in hadronic collisions,
hadronic diffraction, inclusive DIS and 
$e^{+}e^{-}$ annihilation, at the same mass of the hadronic states. 
Non-trivial differences are found when one looks at the fine structure
of final states. 
The thrust analysis reported by ZEUS is
performed on a diffractive event sample selected by the LPS. 
A comparison of these results with those obtained from 
the LRG events in H1 indicates that the average event thrust in the ZEUS data
is systematically higher although, because of the large errors, 
not inconsistent with the H1 findings. 
However, the average 
thrust in the LRG events is definitely smaller than in the $e^+ e^-$ data.

The stumbling block in the interpretation of these data is that
the theoretical understanding of initial and final state radiation 
and of the related virtual radiative corrections to the formation of 
diffractive final states, is lagging behind the rapid experimental development.
The experimentalists have taken the lead and, at the moment, the RAPGAP Monte Carlo,
based on the Ingelman-Schlein approach, remains the only tool to 
describe the resolved Pomeron via partonic densities which 
are obtained from fits to the H1 $x_\Pom .F_{2}^{D(3)}$ structure 
function. The principal finding is that this particular version of RAPGAP 
describes almost all of the 
diffractive hadronic final states ranging from energy flow to particle
correlations. In terms of this model a large gluonic Pomeron content, as
determined from the $x_\Pom .F_{2}^{D(2)}$ analysis, is essential for a good
description of the data, although it should be emphasized that
other Monte Carlo's like  LEPTO~6.5, 
based on the soft colour interaction model
and which does not contain any special mechanism of diffraction
describes the data equally well. 
At present, the data do not allow a discrimination between these 
conceptually different models to be made.

From the theoretical point of view one also should take into account
that  the presently available Monte Carlo models are assuming an illegitimate
Regge factorisation, in which hard scale dependencies on $x_\Pom$ an $\beta$
as found in theoretical QCD analyses, and which characterise the final
state, are neglected. 
For instance, one treats the charm production as entirely due 
to the familiar photon-gluon fusion, neglecting the 
direct charm-anticharm
excitation which some theorists claim to be substantial. 
In this approximation, in order to reproduce the diffractive charm signal
reported by  Thompson~\cite{thompson} 
one needs a hard glue in the Pomeron fits~\cite{whitmore}. 
Therefore the conclusions drawn from these Monte
Carlo studies as to the physical picture underlying the diffractive 
final states should be handled with care.

\section{The Forward Region}

In elastic scattering, the typical impact parameter is a sum
of the size of the target, the projectile and of the range of
interaction between the target and projectile constituents.
In the generic diffractive reaction $ap\rightarrow XY$, the diffractive 
slope $B_{D}$ is close to the slope of elastic hadronic scattering,
$B_{d} \sim 10$ GeV$^{-2}$ in the exclusive limit of small
mass states, $M^2_{X,Y} \approx {\cal {O}} (M_P^2)$,
but the contributions from the $a\rightarrow X$
and $p\rightarrow Y$ transition vertices are known to vanish
as soon as $X$ or $Y$ are high mass continuum states, 
so $B_{D}$ decreases with the increase of  $M_{X}, M_{Y}$. 
By the same token, only the size of the
scattered proton and the interaction range contribute to $B_{D}$
for single diffraction. Hence one expects a universal value for 
$B_{D} \sim $~6-7~GeV$^{-2}$ in single diffraction for all projectiles $a$
into continuum $X$ (including hadrons, $a=p,\pi,K$, as well as
real and virtual photons $a=\gamma, \gamma^*$) in good agreement with 
the observations. The related universality of the $|t|$-dependence is to
be expected at larger $|t|$, and Meng has presented empirical
evidence for that~\cite{meng}. Pronyaev~\cite{pronyaev} has reported
an evaluation of $B_{D}$ for diffractive DIS $ep \rightarrow e'p'X$
in the colour dipole picture of diffraction;
a nontrivial prediction is a substantial rise of the diffraction slope $B_{D}$
from the exclusive limit $\beta\approx 1$, when $X$ is the 1S vector
meson, to excitation of continuum at $\beta \sim 0.5$.

The crucial theoretical point about
leading nucleon production for non-diffractive $z \approx 1 - x_{\Pom} \lsim
0.9$,
and in the fragmentation of protons in general, is that the QCD hardness
scale for secondary particles ($h$) in semi-inclusive
DIS, $e p \rightarrow e' X h$, gradually
decreases from  $Q^{2}$ in the virtual photon (current) fragmentation
region to a soft, hadronic, scale in the proton fragmentation region.
This suggests a similarity between the inclusive spectra of leading baryons
in high energy hadron-proton and virtual photon-proton (DIS) collisions.
The standard QCD hadronization models fail in this manifestly soft part
of the phase space, but were never really meant to describe it.

The non-perturbative mechanisms - pion exchange for the neutron
production and Pomeron+pion+Reggeon exchange for the leading proton
production - have been discussed by D'Alesio~\cite{dalesio} and
Nikolaev~\cite{WG1_nnn}, respectively. As has been understood for many years,
tagging leading neutrons selects DIS off pions. However, the
extraction of the pion structure function at small values of the 
Bjorken variable $x_{\pi}=\beta$ requires the knowledge of the flux of pions.
D'Alesio focused on the model dependence caused by absorption corrections, 
which are different for leading neutron production in hadronic collisions 
and DIS
and spoil the Regge factorization 
leading to an uncertainty of $20-30 \%$ in the associated normalization between processes
(a similar analysis has been reported in~\cite{WG1_nnn}).
The conclusion is that absorption effects are under
reasonable control, and do not preclude the experimental 
determinations of the gross features of the pion structure function.
The related absorption corrections define the Bjorken's gap survival
probability in hard diffractive $pp$ collisions.
The $x,Q^{2}$ evolution properties of the leading
neutron production as reported by Nunnemann for H1~\cite{nunnemann}
and Garfagnini for ZEUS~\cite{garfagnini} are consistent with
expectations of the DGLAP evolution of the pion structure function.

The pQCD-motivated evaluation of Reggeon exchange in diffractive
DIS has been reported by Sch\"afer~\cite{schaefer}.
Reggeon exchange is evaluated in terms of the valence quark
distributions in the proton and comes out at the same order
of magnitude as the H1 evaluations. In this analysis
the strongest possible constructive interference of
the Pomeron $(\Pom)$ and Reggeon $f$ exchanges appears, in contrast to
expectations based on treating the Pomeron and Reggeon as hadronic
states. Furthermore, he showed that the Pomeron, Reggeon, and the
$\Pom f$ interference structure functions must have a similar
large-$\beta$ behaviour and that the $\beta,Q^{2}$ evolution of all these
structure functions must be similar. The latter point leads to an
approximately $x,Q^{2}$-independent yield of leading
protons. Nunnemann~\cite{nunnemann} reported a good agreement of the H1
data on leading protons with the Pomeron+pion+Reggeon exchange
model~\cite{nnn}, whereas LEPTO~6.5 Monte Carlo fails to reproduce
the $Q^{2}$-dependence of the observed cross section.
In principle, Reggeon exchange is constrained by the
diffractive data, but more detailed numerical evaluations of the $\Pom f$
interference are needed for a unified description of the Reggeon 
effects in both the diffractive region $z\gsim 0.9$, and for $z\lsim 0.9$.

\section{Diffraction in proton-proton scattering}

\subsection{New data  from the Tevatron}

Hard diffraction at the Tevatron has been
observed by both the D\O\ (reported by Rubinov~\cite{rubinov})
and CDF (reported by Borras~\cite{borras1}) collaborations.
Diffractive events are selected by requiring a large
rapidity gap and/or by recording a beam particle recoil.
The hard scale is set either by jets with large $E_T$,
or by the mass of a diffractively-produced W-boson (CDF).
In hard single-diffraction with a forward rapidity gap the gap fraction,
defined as the excess of events at low multiplicity over the extrapolated
multiplicity
distribution from non-diffractive dijet events, has been measured.
The dependence of the gap fraction on $\eta_{boost}$ and
the gap location and size indicates that these events are
indeed consistent with a colour singlet production mechanism.
The gap fractions measured by the D\O\ collaboration
at two different energies ($\sqrt{s}=630,1800~GeV$) are of the same
order of magnitude ($ {\cal{O}} (1\%)$~).
The jet transverse energy distribution is similar to those of
non-diffractive event, indicating that the Pomeron has a hard
partonic structure.
Combining the ratio's of diffractive to non-diffractive W and dijet
production, the CDF collaboration determined
a fraction of hard gluons in the Pomeron equal to $0.7\pm0.2$.
This value entails a momentum fraction of the hard partons in the Pomeron
which is consistent with results from the ZEUS experiment only after
introducing a discrepancy factor $D=0.18$~(cf. flux
renormalization).
Both  collaborations also observed events with two central jets
and two gaps in the forward rapidity regions,
consistent with hard double Pomeron exchange.
The rate, R(DPE/SD) = $0.26\pm0.05(stat)\pm0.05(syst.)\%$,
as well as the kinematics of the dijets are well reproduced by Monte Carlo,
provided a renormalized $\Pom$ flux is used.

The fraction of dijet events with a central rapidity gap has been measured
by D\O\ (reported by Goussiou~\cite{goussiou}) and CDF
(reported by Borras~\cite{borras2}) at $\sqrt{s}=630$ 
and $1800$ GeV and is found to 
decrease with the $p\bar{p}$ centre-of-mass energy:
\[
R_{D\O} = \frac{f_{JGJ}(630)}{f_{JGJ}(1800)} = 3.4 \pm 1.3
~~;~~~
R_{CDF} = \frac{f_{JGJ}(630)}{f_{JGJ}(1800)} = 2.0 \pm 0.9 \; .
\]
The gap fraction dependence on the dijet transverse energy and 
pseudorapidity separation shows a
slightly rising (D\O) / rather flat (CDF) tendency,
although the present errors do not allow a clear discrimination.
Various Monte Carlo models for colour-singlet exchange have been fitted 
to the $E_T$ and $\Delta\eta$ dependence of the measured gap fraction 
by the D\O\ collaboration.
Assuming that the survival probability of the gap does not depend on
$E_T$ and $\Delta\eta$, the data favour quark-initiated colour-singlet
processes.

\subsection{A resolution to Dino's paradox ?}

It is well known that the bulk of
elastic and total hadronic cross section data can be
described by the exchange of a soft Pomeron pole with an
intercept greater than one.
Goulianos has pointed out that if one puts this Pomeron into
the triple Regge formula,
which results from factorization of the Regge pole,
and fixes the normalization based on the FNAL-ISR data, then
it overshoots the diffractive $p {\bar p}$
Tevatron data by a factor of 5-10 (this has become known as ``Dino's
paradox''). Clearly the
classical Pomeron flux factor must be modified
in some way to account for this and various
modifications have been suggested. 
Tan~\cite{tan} points out that in the Tevatron data the rapidity gap
is such that the exchanged Pomerons are in a moderate
energy regime. In the conventional Donnachie-Landshoff
fits to total cross section, $\sigma_{tot} =
\sigma_{\Pom}s^{\Delta_{\Pom}}
+\sigma_{R}s^{\Delta_{R}}$, with an $s$-independent $\Delta_{\Pom}$,
the Reggeon exchange contribution is numerically very large.
Tan has emphasized that because in $pp$ scattering there are no $s$-channel
resonances, such a large Reggeon contribution is in conflict with
duality and exchange degeneracy ideas, according to which
$pp$ total cross section must involve pure Pomeron exchange.
Hence, in the moderate energy region relevant to the
rapidity gaps in the Tevatron diffractive data, the
intercept of the Pomeron must be close to unity, which removes
the rapid growth of the triple Regge cross section from the
FNAL-ISR to Tevatron and resolves Dino's paradox. Tan has discussed
flavouring - the effect of opening new
inelastic channels - as the  mechanisms for the energy
dependent intercept of the Pomeron. Tan's mechanism can be confirmed
or ruled out at LHC.

At present, theory is not able to meet the challenge of the extremely
interesting data on hard jet production in rapidity
gap events observed in many jet and $W$-boson
production channels, at the  Tevatron. Bjorken's point that absorption
effects strongly affect the gap survival probability has
been reiterated by Whitmore~\cite{whitmore}, who has presented
evaluations for diffractive jet and $W$-production
for different parton model parameterizations of the H1-ZEUS
diffractive structure functions based on the Regge
factorization assumptions. In all the cases such estimates
overshoot the observed cross sections, which testifies to
the importance of absorption.
Whitmore's results show
that the gap survival probability varies substantially
from one hard diffractive reaction to another, the
theoretical understanding of these variations is, as yet, lacking.

\section{Superhard diffraction and BFKL dynamics}

The evolution with energy (or $1/x$) of the cross section for
scattering of two small objects of similar size,
i.e. Mueller's ``onia'', also called the single-hard-scale problem,
remains one of the most intriguing and difficult problems in perturbative QCD.
Fadin~\cite{fadlip} and Lipatov have presented corrections, next-to-leading in
energy, to their famous BFKL equation; 
these subleading corrections are very large
and reduce the strong rise in energy of hard cross sections, of the
leading-order result.
In view of this, it is vitally important that the experiments continue
their efforts to measure hard small-$x$ processes.

Cox~\cite{cox} and Forshaw have suggested looking at double dissociation
in photoproduction (DDP) at high $|t|$, as an alternative
to the traditional gaps-between-jets measurement.
The latter has several distinct disadvantages: the demand for
high enough $p_t$ in the jets (typically $p_t^2 = 25~$GeV$^2$)
reduces statistics and diminishes the available rapidity space
(the jets themselves occupy as much as two units in rapidity each and must be
seen in the main detectors); one also relies strongly on being
able to measure the size of the gap accurately (which in practice also requires
an experimental definition). In contrast DDP at high $|t|$
which merely uses the gap to separate the two systems $X$ and $Y$
(following the H1  method), has a much wider reach in rapidity (or energy) and
may be relevant to $|t|$ values as low as $1 $~GeV$^2$. Monte Carlo studies,
using HERWIG,
suggest that this measurement is a robust measure of {\it whatever the energy
rise
of the process is}. It will certainly be interesting to see the first data.


\section*{Appendix - Diffractive DIS: Convention Summary}

\section*{Inclusive DIS}
Lorentz-invariant variables 
\begin{eqnarray*}
Q^2 & \equiv & - q^2      =  - (k - k')^2 \nonumber \\
W^2 & \equiv & (p + q)^2  =  M_p^2 + 2p.q - Q^2 \approx 2p.q - Q^2  \nonumber \\
x   & \equiv & \frac{Q^2}{2 p.q }  =  \frac{Q^2}{W^2 + Q^2 - M_p^2} \approx 
\frac{Q^2}{W^2 + Q^2} \nonumber \\
S   & \equiv & (p + k)^2  =  M_p^2 + 2p.k +m_e^2 \approx 2p.k \nonumber \\
y   & \equiv & \frac{2q.p}{2k.p}  =  \frac{W^2 + Q^2 - M_p^2}{S-M_p^2} 
\approx \frac{W^2 + Q^2}{S} \approx \frac{Q^2}{xS} 
\end{eqnarray*}
\section*{Diffractive Processes}
In a general {\bf doubly-dissociative diffractive (DD)} process the final state
consists of fragments of the photon (X) and of the proton (Y) with a large
rapidity gap $\Delta \eta$ (see Figure on page 17).
\begin{itemize} 
\item  Diffractive variables (Eilat convention) 
\begin{eqnarray*}
t & \equiv &  (p - p')^2  =  M_p^2 + M_Y^2 - 2 p.p' \nonumber \\
M_X^2  & \equiv & (p - p' + q)^2 \nonumber \\
x_\Pom & \equiv & \frac{(p-p').q}{p.q} = \frac{M_X^2 + Q^2 - t}{W^2 + Q^2 -
M_p^2}
\approx  \frac{M_X^2 + Q^2 -t}{W^2 + Q^2} \nonumber \\
\beta  & \equiv & \frac{Q^2}{2(p-p').q}  = \frac{x}{x_{\Pom}} =  
\frac{Q^2}{M^2 + Q^2 -t} 
\end{eqnarray*}
\item Pseudorapidity interval between the fragments $X$ and $Y$ 
\begin{displaymath}
\Delta \eta  \approx \log {\frac{1}{x_{\Pom}} \frac{M_Y^2}{M_p^2}}. 
\end{displaymath}
\end{itemize}
A {\bf singly-dissociative diffractive (SD)} process is a special case in which
Y is a proton and
{\begin{center}
{\it $M_Y^2= p'^2= M_p^2$}
\end{center}}
\begin{itemize}
\item At HERA
$t\leq 1~GeV^2$, and can be neglected in the above expressions 
for $\beta,x_{\Pom}$
\item 
$t \approx  - p_{\bot}^{' 2}$ with the transverse plane perpendicular to 
that defined by the incoming ($p,\gamma^*$) in the centre-of-mass frame
\item 
The angular-averaged SD diffractive cross section can be decomposed as
\begin{eqnarray}
\lefteqn{Q^{2}y \frac{d^{4}\sigma(ep \rightarrow ep'X)}{dQ^{2}dy dM_X^{2} dt}
 = } \\
& &  \frac{\alpha_{em}}{\pi}
\left\{ (1-y+ \frac{y^{2}}{2} ) \cdot 
\frac{d^{2}\sigma_{T}(\gamma^{*}p\rightarrow p'X)}{dM_X^{2}dt} 
 +  (1-y) \cdot 
\frac{d^{2}\sigma_{L}(\gamma^{*}p\rightarrow p'X)}{dM_X^{2}dt} \right\} 
\nonumber 
\label{eq:13}
\end{eqnarray}
\item Diffractive structure functions  (H1/ZEUS convention)
\bea
\lefteqn{x_{\Pom} F_{2}^{D(4)}(t,x_{\Pom},\beta,Q^{2}) =} \\
& &\frac{Q^{2}}{4\pi^{2}\alpha_{em}} \cdot
\left\{
\frac{ x_{\Pom} d^{2}\sigma_{T} (\gamma^{*}p\rightarrow hX)}{dx_{\Pom}dt} 
+ \frac{x_{\Pom}d^{2}\sigma_{L}(\gamma^{*}p\rightarrow hX)}{dx_{\Pom}dt}
\right\} \nonumber
\label{eq:15}
\eea
\bea
x_{\Pom} F_{L}^{D(4)}(t,x_{\Pom},\beta,Q^{2})
=
\frac{Q^{2}}{ 4\pi^{2}\alpha_{em}} \cdot
\frac{x_{\Pom}d^{2}\sigma_{L}(\gamma^{*}p\rightarrow hX)}{dx_{\Pom}dt} 
\nonumber
\label{eq:14}
\eea
\item
Parameterizing the t-dependence by the diffractive slope  $B_D$:
\bea
F_{2}^{D(4)}(t,x_{\Pom},x,Q^{2})\approx 
F_{2}^{D(4)}(0,x_{\Pom},x,Q^{2})\exp(B_{D} t) \nonumber
\eea
where  $B_{D}$ can depend on  $x_{\Pom},\beta,Q^{2}$.
\item
The $t$-integrated  diffractive structure functions 
\bea
x_{\Pom} F_{i}^{D(3)}(x_{\Pom},\beta,Q^{2})=
\frac{Q^{2}}{4\pi^{2}\alpha_{em}} 
\int dt \frac{ x_{\Pom} d^{2}\sigma_{i}(\gamma^{*} p \rightarrow Xp')}{dx_{\Pom}dt}
\eea
\item {\bf Exclusive singly-dissociative diffractive (ESD)} (or elastic) vector meson production
is an exclusive limit of SD in which Y is a proton and X is a vector meson,
$M_X=M_\rho,...,M_Y$.
\item {\bf Exclusive doubly-dissociative diffractive processes (EDD)}: X is a vector meson,
$M_X=M_\rho,...,M_Y$ and the proton excites into nucleon resonnances and/or continuum states Y.
\end{itemize}
\begin{figure}[ht]
\begin{center}
\vspace*{-.5cm}
\parbox[b]{11cm}{\psfig{width=10cm,file=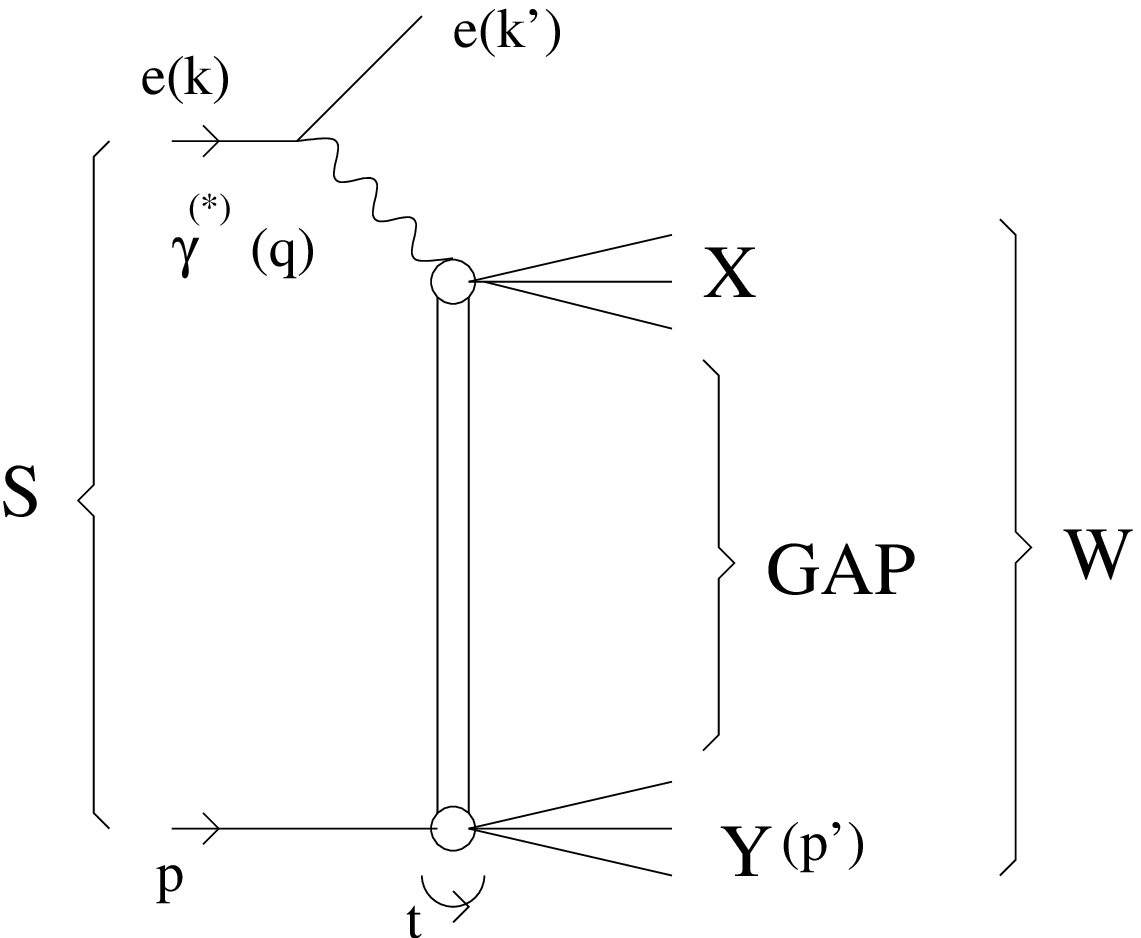}}\\
\end{center}
\refstepcounter{figure}
\label{qq}
\end{figure}

\end{document}